\documentclass[aps,prl,superscriptaddress,a4paper,floatfix,nofootinbib,twocolumn]{revtex4-1}
\usepackage{graphicx,color,amsmath,amssymb,enumerate}%amssymb,
\usepackage[normalem]{ulem}
\usepackage[ocgcolorlinks,colorlinks=true,linkcolor=blue,citecolor=red]{hyperref}
\newcommand{\ra}{\rangle}
\newcommand{\la}{\langle}
\newcommand{\tr}{\textrm{tr}}
\newcommand{\bra}[1]{\langle #1|}
\newcommand{\ket}[1]{|#1\rangle}

\newcommand{\braket}[2]{\langle #1 | #2 \rangle}
\newcommand{\kets}[2]{|#1\rangle_{\!_#2}}

\newcommand{\proj}[1]{|#1\rangle\!\langle#1|}
\newcommand{\op}[2]{|#1\rangle \!\langle#2|}

\newcommand{\order}[1]{\mathcal{O}\!\left( #1 \right)}
\newcommand{\norm}[1]{\left\| #1 \right\|}
\newcommand{\delp}{\delta p}
\newcommand{\delW}{\delta W}
\newcommand{\delF}{\delta F}
\newcommand{\delS}{\delta S}
\newcommand{\delU}{\delta U}
\newcommand{\uz}{\uparrow_z}
\newcommand{\dz}{\downarrow_z}
\newcommand{\beq}{\begin{equation}}
\newcommand{\eeq}{\end{equation}}

\begin{document}
\title{Work extraction and thermodynamics for individual quantum systems}
\author{Paul Skrzypczyk}\affiliation{ICFO-Institut de Ciencies Fotoniques, Mediterranean Technology Park, 08860 Castelldefels, Barcelona, Spain}
\author{Anthony J.~Short}\affiliation{H. H. Wills Physics Laboratory, University of Bristol$\text{,}$ Tyndall Avenue, Bristol, BS8 1TL, United Kingdom}
\author{Sandu Popescu} \affiliation{H. H. Wills Physics Laboratory, University of Bristol$\text{,}$ Tyndall Avenue, Bristol, BS8 1TL, United Kingdom}

\begin{abstract}

Thermodynamics is traditionally concerned with systems comprised of a large number of particles. Here we present a framework for extending thermodynamics to individual quantum systems, including explicitly a thermal bath and work-storage device (essentially a `weight' that can be raised or lowered). We prove that the second law of thermodynamics  holds in our framework, and give a simple protocol to extract the optimal amount of work from the system, equal to its change in free energy. Our results apply to any quantum system in an arbitrary initial state, in particular including non-equilibrium situations. The optimal protocol is essentially reversible, similar to classical Carnot cycles, and indeed, we show that it can be used it to construct a quantum Carnot engine.
\end{abstract}

\maketitle
Thermodynamics forms part of the bedrock of our current understanding of the physical world. It has remained unchanged despite huge revolutions in physics, such as relativity and quantum theory, and few believe it will ever fail. Over time, it has been applied to situations well outside its original domain; from black holes \cite{Bek73,BarCarHaw73}, to quantum engines comprised of only a few qubits \cite{GevKos96,YouMahOba09,TonMah05,LinPopSkr10a,LevKos12}. Drawing inspiration, in part, from the “resource theory” paradigm in quantum information \cite{HorOpp12, BenBerPop96,HorHorHor09, BarRudSpe07, HorHorOpp03, MarSpe11}, recently there has been much renewed interest in the foundations of thermodynamics, with a number of very interesting results already obtained \cite{HorOpp11, BraHorOpp11, BraHorNg13, JanWocZei00,ProLev76, AliHorHor04, GemMicMah04,GyfBer05,HasIshDri10, TakHasDri10, EspVan11, Abe11, VerLac12,AllBalNie04,AllSerNie05,AllJohMah08,AllHov11,DahRenRie11,Abe11,RioAbeRen11,EglDahRen12,FaiDupOpp12,BruLinPop11,Abe13}. One of the overarching fundamental questions that these works are concerned with is of the applicability of thermodynamics to quantum systems; it is this question that we wish to address in this paper.

Thermodynamics was originally invented to deal with macroscopic thermal machines such as steam engines, long before microscopic particles, let alone the theory of quantum mechanics, were discovered. It is therefore plausible that significance differences exist in the quantum regime. Indeed, recent result call into question the role of free energy for individual quantum systems \cite{HorOpp11}. Classical thermodynamics tells us that the total amount of work we are able to extract from a system is given by its change in free energy, which was also supported by previous quantum results \cite{ProLev76,AliHorHor04,HasIshDri10,TakHasDri10,EspVan11,Abe11,VerLac12,AllBalNie04,AllSerNie05,AllJohMah08,AllHov11}. Yet in  \cite{HorOpp11} an alternative paradigm was presented in which it was shown that work equal to free energy can be extracted only if we collectively process many copies of the same system. When acting on each copy individually, the amount of work that can be extracted is generally significantly less than the free energy. Moreover, even more recent results show that considering catalysts \cite{BraHorNg13} further change the story. These results therefore suggest that the free energy is not the relevant quantity for individual systems.

Here we revisit the issue of work extraction and show that free energy is a significant quantity for individual systems.  To do so we present a paradigm for dealing with thermodynamic processes within quantum theory. Our paradigm is similar to that of \cite{HorOpp11}  but differs in two essential aspects. In  \cite{HorOpp11}  they considered almost deterministic work extraction, from the `single-shot' viewpoint which has received much attention lately \cite{DahRenRie11,Abe11,RioAbeRen11,EglDahRen12,FaiDupOpp12,BraHorNg13}. Here, in contrast, we will consider \textit{average} work extraction, and only require \textit{average} energy conservation. In this context, we first prove the second law of thermodynamics holds, and second give a simple protocol which extracts work equal to the free energy change of an individual quantum system and show that this is optimal. We furthermore show that this protocol can be used to construct a quantum Carnot engine similar to the one in \cite{BruLinPop11}, from which our optimality results imply the Carnot limit, an alternative formulation of the second law of thermodynamics. An alternative approach that also allows one to extract average work equal to the free energy change of a system was very recently proposed in \cite{Abe13}, where a key difference is that a re-usable source of coherence is included in the framework.

%\section{Results}\label{s:preliminaries}

\subsection{The paradigm}

In this section, we more precisely describe our framework for quantum thermodynamics. In particular, we define the system, thermal bath, and work-storage device, and give explicit definitions for thermodynamic quantities such as heat, work, free energy, and entropy within our framework. In light of this, we consider the allowed transformations, and impose the first law of thermodynamics.

We consider any quantum system (of finite-dimension), in an arbitrary initial state $\rho_\mathrm{S}$ and with arbitrary Hamiltonian $H_\mathrm{S}$. In accordance with statistical mechanics, we define the system's internal energy as $U=\tr(\rho_\mathrm{S} H_\mathrm{S})$ (i.e. its average energy), and its entropy as the  von Neumann entropy $S = - \tr (\rho_\mathrm{S} \log \rho_\mathrm{S})$.  Note that the system itself need not have a well-defined temperature, however, its free energy relative to a thermal bath at temperature $T$ is given by $F=U-TS$.

To represent a thermal bath at temperature $T$, we assume that  we have an unlimited supply of finite-dimensional systems, each with any desired Hamiltonian  $H_\mathrm{B}$,  in the corresponding thermal state $\tau_\mathrm{B} = \frac{1}{\mathcal{Z}} \exp\left(-\frac{H_\mathrm{B}}{T}\right)$, where $\mathcal{Z}~=~\tr\left(\exp\left(-\frac{H_\mathrm{B}}{T}\right)\right)$ is the partition function, and we set $k_B=1$ throughout for convenience. When one has access to a thermal bath, any system in a thermal state is essentially a `free resource' \cite{BraHorOpp11}. Note that any physical protocol must involve a finite number of  systems from the bath, which can be thought of as a single large thermal system. We define the heat flow $Q$ out of the bath as the decrease in its average energy, i.e. if the bath system is transformed into state $\sigma_\mathrm{B}$, then $Q=\tr(H_\mathrm{B} (\tau_\mathrm{B} -\sigma_\mathrm{B}))$.

In this work we wish to explicitly include the physical device which stores the work we extract. The work storage device we will consider here is a suspended weight, which is raised or lowered when work is done on or by it. In particular, we will consider a quantum system whose height is given by the position operator $\hat{x}$, with Hamiltonian $H_\mathrm{W}= m g \hat{x}$ representing its gravitational potential energy.  For simplicity, we choose $mg =1 \textrm{ Jm}^{-1}$, such that the value of  $\hat{x}$ directly denotes the work stored by the mass.  Such a system has a long history of being used as a work storage system in classical thermodynamics \cite{LieYng99}. %also used by Johan

We define the work $W$ extracted  as the change in the average energy of the weight. Hence if the weight is initially in the state $\rho_\mathrm{W}$ and is left in the state $\sigma_\mathrm{W}$, then $W=\tr(H_\mathrm{W} (\sigma_\mathrm{W} - \rho_\mathrm{W}))$. We do not place any constraints on the initial state of the weight, unlike in \cite{Abe13}. In fact, as we will see below, by construction the explicit choice of initial state will play no role in this work.

It will be helpful to define the translation operator $\Gamma_a$, which acts on (un-normalised) position states of the weight as $\Gamma_a\ket{x} = \ket{x+a}$.

In previous work \cite{HorOpp11} an alternative work-storage system was suggested -- raising a qubit deterministically from its ground state to it's excited state. This qubit was termed a wit, short for work bit. However, choosing the energy gap of the work bit requires advance knowledge of the work to be extracted, and so this model does not translate well to non-deterministic work extraction, which we will be interested in here. Furthermore, we would prefer to be able to use a single work storage system as a `battery' capable of gaining and expending work in multiple thermodynamic processes.

We assume that the initial state is a product state of the system, bath and weight. We now consider the allowed transformations in our framework. The intention here is to remain as general as possible, whilst eliminating the possibility of `cheating' by bringing in resources from outside the framework (such as external sources of work or free energy), or making use of objects within the framework for a purpose other than intended (for example, by using the work-storage device as a cold reservoir in a heat engine). Our first two assumptions are very general: 
The first is unitarity. The most general quantum transformation is a completely positive trace-preserving map. However, here we consider only unitary transformations of the system, bath and weight. This prevents us from using external ancillas in non-thermal states as a source of free energy.
The second is average energy conservation (the first law). We require that any particular protocol conserves the total average energy  (for the particular initial state of the system and bath on which it is designed to operate, and on any initial state of the weight). In terms of the quantities defined earlier, this corresponds to the first law of thermodynamics, which with our chosen sign convention can be expressed as
\begin{equation} \label{e:1stlaw}
\Delta U  = Q - W.
\end{equation}
This prevents us from using the transformation itself as a source of work (for example, by simply raising the weight). Note that this assumption differs from that made in previous works \cite{BraHorOpp11,HorOpp12} that the unitary evolution commutes with the total energy operator. We will comment more on this in the Conclusions.

We also place two additional constraints on the allowed dynamics governing interactions with the weight:\renewcommand{\labelenumi}{(\roman{enumi})}
The first is weight-state independence,  that the  work extracted in an allowed protocol must be independent of the initial state of the weight.  Intuitively this is because we want to weight to play a 'passive' role, such that its sole purpose is to keep account of the extracted work. More importantly though this prevents us from `cheating' by using the weight for purposes other than as a work storage system (e.g. as a cold reservoir, or a source of coherence). Furthermore, this ensures that we can use the same work storage system for multiple thermodynamics protocols (or on several copies of the same state) without having to worry how its initial state has been modified by earlier procedures. We prove that our protocol obeys this assumption in the Appendix. \label{i:I}
Finally we demand weight-translation invariance, that any allowed unitaries to commute with translation operations on the weight. This reflects the translational symmetry of the weight system, and the fact that only displacements in its height are important.

\subsection{The second law} \label{s:second_law}
We now show that the second law of thermodynamics holds in our  framework, by proving that there is no protocol which extracts a positive quantity of  work from a thermal bath whilst leaving the system unchanged (i.e. that there is no way of turning heat into work) \cite{LadPreSho08}. To show this we will use proof by contradiction.

Consider a thermal bath at temperature $T$,  an arbitrary quantum system (acting as a working system for the protocol), and a weight.

Let us first consider the energy changes during the protocol. As the final state of the system is the same as its initial state, its average energy cannot change. Suppose that we are able to extract average energy from the bath and store it in the weight, $\Delta E_\mathrm{W} > 0$. The average energy of the thermal bath must change by $\Delta E_\mathrm{B} = - \Delta E_\mathrm{W} $ due to average energy conservation.

Now consider the entropy changes during the same protocol (in particular, the changes in von Neumann entropy $S(\rho) = - \tr (\rho \log \rho))$. As the system, bath and weight are initially uncorrelated, their initial entropy is simply the sum of their individual entropies.  Unitary transformations conserve the total entropy, $\Delta S_\mathrm{total} = 0$. However, as correlations can form during the protocol, the sum of the final entropies can be greater than the sum of their initial entropies (as the entropy is subadditive). This means that
\begin{equation}\label{e:entropy eqn}
	\Delta S_\mathrm{B} + \Delta S_\mathrm{W} + \Delta S_\mathrm{S} \geq \Delta S_\mathrm{total} = 0
\end{equation}

As the final state of the system must be the same as its initial state $\Delta S_\mathrm{S}=0$. Furthermore, given an initial thermal state for the bath (with positive temperature), any change of the state which reduces its average energy must also reduce its entropy (since the thermal state is the maximal entropy state with given average energy), $\Delta S_\mathrm{B} < 0$. However, within our framework all allowed protocols are such that the  work extracted is independent of the initial state of the weight; we are therefore free to choose any initial state of the weight we like. We show in the Appendix that the entropy change of the weight can be made as small as desired by taking its initial state to be a very broad wavepacket (with well-defined momentum). In particular, we can make $\Delta S_\mathrm{W} < |\Delta S_\mathrm{B}|$. This would result in violating \eqref{e:entropy eqn}. Hence there is a contradiction, and thus there is no way to extract work from the bath.

The second law places an upper bound on the amount of work that can be extracted from a system. In the following section, we will show that we can come as close as desired to extracting this maximum amount of work, by presenting an explicit protocol.

\subsection{Extracting work from an individual quantum system}\label{s:work}
Our protocol for extracting work from a quantum system  proceeds in two stages. In the first stage, we  transform the state of the system into a mixture of energy eigenstates, without using the thermal bath. In the second stage, we gradually  transform the system into a thermal state in a sequence of steps, each of which involves a new  qubit from the  bath.  Each step is essentially an infinitesimal Carnot cycle, similar to the one discussed in \cite{BruLinPop11}. In both stages we extract an amount of work arbitrarily close to the free energy change of the system. It follows from our proof of the second law that this protocol is optimal.

Stage 1 is to transform the system into a mixture of energy eigenstates. In this stage we transform the system into a mixture of energy eigenstates without using the thermal bath, and extract work equal to its change in free energy. Consider a system and  weight, initially in an arbitrary uncorrelated state represented by the density operator $\rho_\mathrm{S} \otimes \rho_\mathrm{W}$.

We can always expand $\rho_\mathrm{S}$ in terms of its eigenvalues $p_n$ and  eigenvectors $\ket{\psi_n}$ as $\rho_\mathrm{S} = \sum_n  p_n \proj{\psi_n}$,
where we have ordered the eigenvalues such that $p_{n+1} \leq p_n$. Denoting the energy eigenstates of the system by $\ket{E_n}$ (with corresponding eigenvalues $E_n$), we implement the unitary transformation
\begin{equation}
V=\sum_n \op{E_n}{\psi_n} \otimes \Gamma_{\varepsilon_n},
\end{equation}
where $\epsilon_n =\bra{\psi_n} H_\mathrm{S} \ket{\psi_n} - E_n$, such that $V$ always conserves average energy. After the transformation, the final state is
\begin{equation}
\sigma_\mathrm{SW} = \sum_n p_n \proj{E_n} \otimes \Gamma_{\epsilon_n}\rho_\mathrm{W}  \Gamma_{\epsilon_n}^{\dagger}
\end{equation}
with the reduced states $\sigma_\mathrm{S} =\tr_\mathrm{W} (\sigma_\mathrm{SW})$ and $\sigma_\mathrm{W} =\tr_\mathrm{S} (\sigma_\mathrm{SW})$. The work extracted is given by $W~=~\tr \left( \sigma_\mathrm{W} H_\mathrm{W} \right) ~-~\tr_\mathrm{W}(\rho_\mathrm{W} H_\mathrm{W}) ~=~\sum_n p_n \epsilon_n$.
The change in average energy of the system is $\Delta U ~=~\tr \left( \sigma_{S} H_\mathrm{S} \right) ~- ~\tr_\mathrm{S}(\rho_\mathrm{S} H_\mathrm{S}) ~= ~-~\sum_n p_n \epsilon_n$,
 hence this protocol is in accordance with  the first law of thermodynamics (i.e. $\Delta U = - W $). Futhermore the entropy of the system remains unchanged, so the work  extracted is precisely equal to the free energy lost by the system
\begin{equation}
\Delta F = F(\rho_\mathrm{S}) - F(\sigma_\mathrm{S})  = \Delta U =  -W.
\end{equation}
Stage 2 of the protocol consists in extracting work from a mixture of energy eigenstates. In this stage we show that it is possible to extract work equal to the free energy change when transforming a system between two states which are diagonal in the energy eigenbasis. By transforming the state $\sigma_\mathrm{S}$ obtained in the first stage into a thermal state, we extract the maximum amount of work from the system.

We begin by considering a small change in the occupation probabilities of  two energy levels. In particular, suppose that we wish to transform the state $\sigma_\mathrm{S} = \sum_n p_n \proj{E_n}$ into a new state $\sigma_\mathrm{S}' =\sum_n q_n \proj{E_n}$, in which $q_1 = p_1+ \delp$,  $q_0 =p_0- \delp$, and $q_k=p_k$ in all other cases (i.e. for all $k>1$).  We consider the situation in which $|\delp| \ll p_1 \leq p_0$. Note that this excludes the case in which $p_1=0$, which introduces some additional subtleties that are detailed in the Appendix. Nevertheless the protocol and conclusions presented below are unchanged.

To achieve the above transformation, we take a qubit from the thermal bath with energy  eigenstates $\kets{0}{\mathrm{B}}$ and $\kets{1}{\mathrm{B}}$ such that its state has the form
\begin{equation} \label{e:bath_state}
\rho_\mathrm{B} = \frac{q_0}{q_0+q_1} \proj{0} + \frac{q_1}{q_0+q_1} \proj{1}
\end{equation}
 i.e. such that the ratio of ground and excited state populations matches that of the corresponding states in $\sigma_\mathrm{S}'$.
Note that this fixes the energy spacing $E_\mathrm{B}$ of the qubit, as $q_1= q_0 \exp \left(-\frac{E_\mathrm{B}}{T} \right)$, hence $E_\mathrm{B} = T \log \left(\frac{q_0}{q_1} \right)$. We then apply the unitary transformation which swaps the bath qubit with the state of the system if the system is in the  two-dimensional subspace spanned by  $\kets{E_0}{\mathrm{S}}$ and $\kets{E_1}{\mathrm{S}}$, and translates the weight to conserve the total energy.  This transformation maps
\begin{equation} \label{e:U2}
	\kets{E_0}{\mathrm{S}}\kets{1}{\mathrm{B}}\kets{x}{\mathrm{W}} \leftrightarrow \kets{E_1}{\mathrm{S}}\kets{0}{\mathrm{B}}\kets{x+\epsilon}{\mathrm{W}}
\end{equation}
for all $x$, where $\epsilon =E_\mathrm{B} - (E_1 -E_0) $,  whilst leaving all other states invariant. This leaves the system in the state $\sigma_\mathrm{S}'$ (For more details of this see the Appendix). Note that this unitary commutes with the  Hamiltonian $H=H_\mathrm{S} + H_\mathrm{B} + H_\mathrm{W}$, so it will obey the first law of thermodynamics.

As the weight is only shifted up or down by $\epsilon$ when the system and bath are in $ \kets{E_0}{\mathrm{S}}\kets{1}{\mathrm{B}}$ or $\kets{E_1}{\mathrm{S}}\kets{0}{\mathrm{B}}$  respectively,  the work extracted is given by $\delW =  \epsilon \delp$. The change in the free energy of the system is given by $\delF = \delU - T \delS$. As we show in the the Appendix, it is straightforward to see that $\delU =  \delp (E_1-E_0)$, and that the change in the entropy of the system is given by $\delS =  \delp \frac{E_\mathrm{B}}{T} + \order{\delp^2}$. Hence to first order in $\delp$, $\delF \approx  \delp (E_1-E_0-E_\mathrm{B}) = - \delp \epsilon = - W$. This shows that we extract work equal to the reduction in free energy of the system, up to a deficit of $\order{\delp^2}$.

In order to extract the maximal amount of work from a quantum state, we perform a sequence of $N$ steps like the one above, interacting the system with a new thermal qubit in each step,  until the system has been transformed into a thermal state at temperature $T$. In particular, for sufficiently large $N$, we can choose a sequence of  $N+1$ states for the system in which subsequent states only differ by a transfer of probability $\delp = \order{\frac{1}{N}}$ between a pair of energy levels, with the first and last states equal to the initial state of the system and its thermal state respectively (for example, starting at the highest energy level, we could first shift probability from all energy levels with higher probability in $\sigma_\mathrm{S}$ than in $\tau_\mathrm{S}$ to the $\kets{E_0}{\mathrm{S}}$ state, then move probability from  $\kets{E_0}{\mathrm{S}}$  to the remaining levels, using $N/(d-1)$ steps for each pair of levels). Applying a unitary of the form \eqref{e:U2} in each step, the work extracted from this stage of the protocol will be
\begin{equation} \label{e:work_deficit}
W= F (\sigma_\mathrm{S}) - F(\tau_\mathrm{S})  - \order{\frac{1}{N}}
\end{equation}
In the limit  $N \rightarrow \infty$ the work extracted will equal the free energy change of the system, regardless of the precise choice of path. Note that in the limit of large $N$, the state of each thermal qubit is only changed slightly by the protocol.

Next we move on to the question of reversibility and optimality of the protocol. By combining both stages of the protocol, and using a sufficiently large number of thermal qubits, it is clear that we can transform an arbitrary state $\rho_\mathrm{S}$ into a thermal state $\tau_\mathrm{S}$ and extract an amount of work as close as we like to the free energy change of the system. The limiting amount of work we can achieve is therefore $W_\mathrm{max}= F (\rho_\mathrm{S}) - F(\tau_\mathrm{S})$.

Interestingly, if $\rho_\mathrm{S}$ is full-rank (i.e. it has no zero eigenvalues), we can also implement the reverse process to create  $\rho_\mathrm{S}$ from an initially thermal system taken from the bath. We  can use the stage 2 protocol to move from $\tau_\mathrm{S}$ to $\sigma_\mathrm{S}$, and then apply the  inverse of the stage 1 transformation. The work cost for this will be  $ W= F (\tau_\mathrm{S}) - F(\rho_\mathrm{S})$. Note that it is not possible to use our protocol to create a state which is not full-rank, as the final step would require the use of a thermal qubit with $E_\mathrm{B}=\infty$. However, as there are always full-rank states arbitrarily close to every state this is not a physically significant restriction. In this sense all transitions between states can be implemented in a thermodynamically reversible way (we note however that if a state is thermalised and then recreated using our protocol, the fluctuations in the position of the weight will increase). This differs from the results of \cite{Abe11,HorOpp11}, who show that such transitions are irreversible when considering  (almost) deterministic work extraction, rather than average work.  Similarly, an arbitrary transformation of the system from a state $\rho_\mathrm{S}$ to $\rho_\mathrm{S}'$ can be achieved (when $\rho_\mathrm{S}'$ is full-rank) for a work cost as close as desired to the free energy change of the system; One way this could be achieved is to transform the system into a thermal state, and then transform the thermal state into the final state.
We now show that our protocol is optimal, using our proof of the second law (See Appendix).
Suppose that there exists a different protocol which extracts  work $F(\rho_\mathrm{S}) - F_\beta(\tau_\mathrm{S}) + \delta$ (where $\delta > 0$) when the system is transformed from $\rho_\mathrm{S}$ to $\tau_\mathrm{S}$. We can then use the above protocol to return the state from $\tau_\mathrm{S}$ to $\rho_\mathrm{S}$, extracting work   $F(\tau_\mathrm{S}) - F_\beta(\rho_\mathrm{S}) - \epsilon$, where we choose the number of thermal qubits  such that $ \epsilon$  is in the range $0 < \epsilon < \delta/2$. The net effect is that a positive amount of work $\geq \delta/2$ is extracted, and the system begins and ends the combined procedure in the same state $\rho_\mathrm{S}$, in violation of the second law.

\subsection{A quantum Carnot engine} \label{s:Carnot}
In previous work \cite{BruLinPop11} a quantum Carnot engine was described, and an argument was made that essentially all Carnot engines are the same. Indeed, each infinitesimal step of  stage 2 of our protocol is essentially the action of such a Carnot engine (although  the situation is more complicated here as the engine has to adapt between steps).  Moreover the frameworks are very different -- Hamiltonian versus unitary, master equations versus extracting qubits from the bath, etc. It is therefore essential to verify that in our present framework we can implement a full Carnot engine. In this section, we show that this is indeed the case.

We must now consider two thermal baths, a hot bath with temperature $T_\mathrm{H}$, and a cold bath with temperature $T_\mathrm{C} < T_\mathrm{H}$. As before, we also have a quantum system (used as a working system that links the two baths) and a weight. Imagine that the system is initially in the thermal state relative to the cold bath (with internal energy $U_\mathrm{C}$ and entropy $S_\mathrm{C}$). Our Carnot cycle is as follows:
First bring the system into contact with the hot bath and use the protocol given in the previous subsection to transform it into the thermal state at temperature $T_\mathrm{H}$ (with internal energy $U_\mathrm{H}$ and entropy $S_\mathrm{H}$). In the asymptotic limit, this allows us to extract work equal to the free energy change of the system with respect to the hot bath. Second, move the system back into contact with the cold bath and use the same protocol to transform it back into the thermal state at temperature $T_\mathrm{C}$, extracting work now equal to the free energy change with respect to the cold bath. 
In the Appendix we show that the total work extracted in both steps is $W = (T_\mathrm{H} - T_\mathrm{C})(S_\mathrm{H} - S_\mathrm{C})$. Furthermore, by applying the first law of thermodynamics to the first step it follows that $Q_\mathrm{H} = T_\mathrm{H} (S_\mathrm{H} - S_\mathrm{C})$. Therefore, combining these two results yields
\begin{equation}
\frac{W}{Q_\mathrm{H}}  = 1- \frac{T_\mathrm{C}}{T_\mathrm{H}}
\end{equation}
which is precisely the Carnot efficiency.  By running this process backwards, we can also construct the corresponding heat pump.

As in standard thermodynamics, the second law prevents us from constructing any heat engine more efficient than the one above. If such an engine were possible, we could subsequently run our Carnot engine as a heat pump such that the net heat flow into the cold bath was zero. In this case, work would be extracted and the hot bath would decrease in energy by a finite amount. The entropy of the hot bath must have decreased as a result, and the entropy of the cold bath can also only have decreased (as it's average energy is unchanged, and it was originally in a thermal state). As before,  the entropy increase of the weight can be made as small as you like by choosing an appropriate initial state. This creates a contradiction with the total entropy conservation expressed in (\ref{e:entropy eqn}).

\section{Conclusions}\label{s:conclusions}
In summary, in this paper we presented a framework for extending thermodynamics to individual quantum systems. Within this framework we proved that the second law of thermodynamics holds and gave an explicit protocol to extract the maximum amount of work from an arbitrary individual quantum system in conjunction with a thermal bath. This work is equal to the change in free energy of the system. Our results apply to any quantum system in an arbitrary initial state, in particular including non-equilibrium situations.  The optimal protocol is essentially reversible, similar to classical Carnot cycles, and indeed, we can use it to construct a quantum Carnot engine.

A key element of our framework is to associate classical energetic quantities (internal energy, heat and work) to average quantum quantities.  However, we want to emphasize that although we use average quantities, we do not require an ensemble of multiple quantum systems to be processed collectively; our protocols act on an individual quantum system.

A significant difference between our framework and other approaches \cite{HorOpp11, BraHorOpp11, BraHorNg13, JanWocZei00} is that the allowed transformations need only satisfy  {\it average} energy conservation  rather than the stronger requirement of strict energy conservation (i.e. unitaries that commute with the total Hamiltonian).  The fundamental reason for allowing such transformations is that average energy conservation corresponds precisely to the first law of thermodynamics in our framework (as we defined all energetic quantities in terms of averages). Demanding strict energy conservation is more than the first law (in our framework) requires.

Allowing protocols that only conserve average energy has major consequences. In particular, when the initial state of the system contains coherences between energy levels, protocols satisfying strict energy conservation cannot extract work equal to the full change in free energy (see Appendix). The work “deficit” equals the difference in free energy between the true state and its energy decohered  version. That is, such protocols simply cannot make any use of the free energy associated to coherences between energy levels. In our protocol this free energy is extracted in stage 1, which is the only part of the protocol that does not satisfy strict energy conservation.

It is interesting to note a further subtle difference between average energy conserving unitaries and strict energy conserving ones. 
In order to be optimal both must be ``state dependent'', i.e. they have to be designed with a particular initial state of the system in mind. However, if we use a unitary designed for a particular system state on a different initial state, in  the  case of strict energy conservation, the external machinery that implements the unitary continues to remain ``neutral", i.e. it doesn't change the energy of the system-bath-weight complex, while in the average energy conserving case the external machinery may exchange energy with the complex.

We also note that unlike in classical thermodynamics, there will also be fluctuations, for example in the final position of the weight. Analysing these fluctuations is an interesting issue for future study.

A subtle aspect that we want to mention is that as our protocols involve a sequence of changing unitaries, we have implicitly assumed the existence of an external clock by which to control the protocol. This raises some interesting points –- the first is whether the clock is a resource which costs work to establish and maintain (in which case we may have over-estimated the amount of work we can extract). Second is to extend the framework to explicitly incorporate the clock, with protocols being implemented via a global time-independent Hamiltonian. Finally in the light of the difference between strictly energy conserving unitaries and average energy conserving unitaries, it is important to investigate whether or not there is any essential difference in the use they make of the energy coherence in the clocks. These are very interesting areas for future work.

Recently there has been considerable progress in studying and understanding the foundations of statistical mechanics, see for example \cite{PopShoWin06,Rei08,LinPopSho09,Rei10,Sho11,RieGogEis12}. It would be extremely important to connect the present results on quantum thermodynamics to that line of research.

To conclude, the resource theory framework seems to be a natural and powerful way to approach thermodynamics. It has already delivered significant results and we believe that further investigation along these lines will lead to a much deeper understanding of the foundations of thermodynamics.

\section{Acknowledgments}  SP acknowledges support from the European project (Integrated Project ``Q-ESSENCE"), the ERC (Advanced Grant ``NLST'') and the Templeton Foundation. AJS acknowledges support from the Royal Society. PS acknowledges support by the Marie Curie COFUND action through the ICFOnest program. PS is grateful to Lidia del Rio, Johan {\AA}berg, Joe Renes, Philippe Faist, Renato Renner, Oscar Dahlsten and Lluis Masanes for interesting discussions.

\section{Appendices}
\subsection{Independence of the work on the initial state of the weight} \label{a:work_independence}

Each step of our protocol can be represented by a unitary transformation of the form
\begin{equation}
V = \sum_{i}  \ket{i}\bra{\tilde{i}} \otimes \Gamma_{a_{i}}
\end{equation}
where $a_i = \bra{\tilde{i}} H_\mathrm{SB}\ket{\tilde{i}}-\bra{i}H_\mathrm{SB}\ket{i}$, the states $ \ket{i}$ and $\ket{\tilde{i}}$ form different orthonormal bases for the system and the relevant portion of the bath, and $\Gamma_a$ is the translation operator on the weight, given by $\Gamma_a=\exp(-i a \hat{p} / \hbar)$, where $\hat{p}$ is the usual momentum operator satisfying  $[\hat{x}, \hat{p}] = i \hbar$.

It is easy to see that all such unitaries commute with translations on the weight. We now show that the  work extracted by a unitary of this form does not depend on the initial state of the weight (even if it is initially correlated with the state of the system).  Let us denote the initial state of the system, bath and weight by the density operator $\rho$. The  work extracted is given by
\begin{eqnarray}
W &=& \tr \left( H_\mathrm{W}  V \rho V^{\dag} \right) -  \tr \left( H_\mathrm{W} \rho \right), \nonumber \\
&=& \tr \left(  \left(V^{\dag} H_\mathrm{W}  V - H_\mathrm{W} \right) \rho \right) \label{eq:workindependence}
\end{eqnarray}
where $H_\mathrm{W} = \openone_\mathrm{SB} \otimes \hat{x}_\mathrm{W} $ is the Hamiltonian of the weight (defined for convenience as an operator on the entire system). Now note that
\begin{eqnarray}
V^{\dag}  H_\mathrm{W} V  &=& \sum_{ij}  \ket{i}\braket{\tilde{i}}{\tilde{j}}\bra{j} \otimes \Gamma_{-a_i} \hat{x} \Gamma_{a_j}  \nonumber \\
&=& \sum_{i} \proj{i} \otimes  \Gamma_{-a_i} \hat{x} \Gamma_{a_i} \nonumber \\
&=& \sum_{i} \proj{i} \otimes  \left(\hat{x} + a_i \openone \right)  \nonumber \\
&=& H_\mathrm{W} + \sum_{i} a_i \proj{i} \otimes \openone
\end{eqnarray}
 Inserting this expression in
(\ref{eq:workindependence}) and simplifying, we obtain
\begin{equation}
W = \sum_{i} a_i \bra{i}   \rho_\mathrm{SB} \ket{i}
\end{equation}
where $\rho_\mathrm{SB} = \tr_\mathrm{W} (\rho)$ is the reduced density matrix of the system and bath. Hence the  amount of work extracted is independent of the initial state of the weight as desired.

\subsection{Proof of second law of thermodynamics}  \label{a:second_law}
In this section, we provide further details for our proof of the second law.

We first argue that any reduction in the average energy of an initially thermal state (with positive temperature) must also yield a reduction in its entropy. We will only need to use the fact that a thermal state is the maximal entropy state with a given average energy, which is well-known and can easily be proven using the method of Lagrange multipliers to maximize the entropy subject to the constraints that the average energy is constant and the state is normalised. If a system starts in a thermal state and is transformed to a final state with fixed average energy, the entropy change $\Delta S_\mathrm{B}$ will be maximised when the final state is also thermal. In the case where the average energy decreases, and the initial state has positive temperature, the final state will be a thermal state of lower temperature, and thus lower entropy.

We now show that the entropy change of the weight in any particular protocol may be made as small as desired by choosing its initial state to be a very narrow wavepacket in momentum space (corresponding to a very broad wavepacket in real space).

As any allowed unitary transformation must  be  invariant under translations of the weight, it can always be written as
\begin{equation}
V = \int \, \textrm{d}p\left(\sum_{ij} v_{ij}(p) \op{i}{j} \right) \otimes \proj{p}
\end{equation}
where the first element of the tensor product corresponds to a unitary operation on the combined system and bath (as a function of the weight momentum) and the second corresponds to a projector onto the un-normalised momentum eigenstate $\ket{p}$ of the weight.  We choose the  basis states $\ket{j}$  to be the eigenbasis of the initial system and bath state, so $\rho_\mathrm{SB} = \sum_j \lambda_j \proj{j}$.

For $V$ to be well defined, there must exist a momentum $p_0$ at which $v_{ij}(p)$ is a continuous function of $p$ for all $i,j$. Define new functions $\eta_{ij}(p) \equiv v_{ij}(p) - v_{ij}(p_0)$ corresponding to the small variations in $v_{ij}(p)$ about $p_0$.

 For any $\epsilon>0$, we can construct an initial pure state  of the weight $\ket{\phi}$ which has support  on a sufficiently narrow interval in momentum space (centred on $p_0$), such   that $|\eta_{ij}(p)| \leq \epsilon$ for all $i,j$ whenever $\phi (p) \equiv \braket{p}{\phi}\neq 0$.

The final state of the weight is given by
\begin{align}
\rho_\mathrm{W}' &= \tr_\mathrm{SB} \left( V \left( \sum_j \lambda_j \proj{j} \otimes \proj{\phi}\right) V^{\dagger} \right) \nonumber \\
&=\sum_{ij} \lambda_j \int\! \textrm{d}p\!\int \textrm{d}q\, v_{ij}(p) \phi(p)  \phi^*(q) v^*_{ij}(q) \,\op{p}{q}  \nonumber \\
&= \proj{\phi} + \sum_{ij} \lambda_j \bigg( v_{ij}^*(p_0)\op{\tilde{\xi}_{ij}}{\phi} \nonumber \\
&\quad\quad\quad\quad+ v_{ij}(p_0)\op{\phi}{\tilde{\xi}_{ij}} + \op{\tilde{\xi}_{ij}}{\tilde{\xi}_{ij}} \bigg),
\end{align}
where  $\ket{\tilde{\xi}_{ij}}$ is the un-normalised state (with norm $\leq \epsilon$):
\begin{equation}
\ket{\tilde{\xi}_{ij}} = \int\! \textrm{d}p\, \eta_{ij}(p)  \phi(p)  \ket{p}
\end{equation}
The  distance between the initial and final states of the weight in terms of the trace norm, defined as $\norm{M}  = \tr \sqrt{M^{\dagger} M}$, is
\begin{align}
\norm{\rho_\mathrm{W}'\!-\! \rho_\mathrm{W}}  &\leq \sum_{ij} \lambda_j  \left(\norm{ \op{\tilde{\xi}_{ij}}{\phi}} + \norm{\op{\phi}{\tilde{\xi}_{ij}}} + \norm{\op{\tilde{\xi}_{ij}}{\tilde{\xi}_{ij}}} \right), \nonumber \\
& \leq d ( 2 \epsilon + \epsilon^2),
\end{align}
where $d$ is the combined dimension of the system and bath (note that this only includes the finite number of systems from the bath used in the  protocol).

As the final state of the weight lives in a finite dimensional subspace, its entropy can be shown to be continuous due to Fannes' inequality \cite{NieChu00}.
\begin{equation}
|S(\rho_\mathrm{W}) - S(\proj{\psi_L})| \leq D \log \left( \frac{d^2}{D} \right)
\end{equation}
where $ D = \frac{1}{2} \norm{\rho_\mathrm{W}'\!-\! \rho_\mathrm{W}} $ is the trace distance between the initial and final states of the weight. We can make $ D$ as small as we like by choosing sufficiently small $\epsilon$ and therefore make the entropy change as small as we like.

\subsection{Work extraction details} \label{a:work_extraction}
In this section we  provide further details regarding stage 2 of our protocol.

We begin by showing that the final state of the  system is  $\sigma_\mathrm{S}'$ after applying the protocol, where
\begin{equation}
\sigma_\mathrm{S}' =\sum_n q_n \proj{E_n}
\end{equation}
is such that $q_1 = p_1+ \delp$,  $q_0 =p_0- \delp$ and $q_k=p_k$ in all other cases (i.e. for all $k>1$).  At the beginning of stage 2 of the protocol, the combined state of the system, bath and weight can be written as
\begin{equation} \label{e:rho}
\rho = \frac{1}{q_0+q_1} \sum_n p_n \proj{E_n} \otimes \left(q_0 \proj{0} + q_1 \proj{1} \right) \otimes \rho_\mathrm{W}^{(n)}.
\end{equation}
where $\rho_\mathrm{W}^{(n)} = \Gamma_{\epsilon_n}\rho_\mathrm{W}  \Gamma_{\epsilon_n}^{\dagger}$. After applying the unitary $V$ which is given by (\ref{e:U2}), and can also be expressed in terms of the translation operator $\Gamma_a$ as $V =\openone \otimes \openone \otimes \openone  +\op{E_1}{E_0} \otimes \op{0}{1}  \otimes\Gamma_{\epsilon }+ \op{E_0}{E_1} \otimes\op{1}{0} \otimes \Gamma_{-\epsilon} - \op{E_0}{E_0} \otimes \op{1}{1}  \otimes\openone - \op{E_1}{E_1} \otimes \op{0}{0} \otimes \openone $, we find
\begin{align}
\rho_\mathrm{S}S' &=  \tr_{BW} \left( V \rho  V^{\dag} \right) \nonumber \\
&=\left( \frac{q_0 p_0 + q_0 p_1}{q_0+q_1} \right)  \proj{E_0} + \left( \frac{q_1 p_0 + q_1 p_1}{q_0+q_1} \right)  \proj{E_1}  \nonumber \\ &\qquad +\sum_{n \geq 2} p_n \proj{E_n} \nonumber \\
&=q_0 \proj{E_0} + q_1 \proj{E_1} +\sum_{n \geq 2} p_n \proj{E_n} \nonumber \\
&= \sigma_\mathrm{S}'
 \end{align}
In later steps of the protocol, the state can be written in a similar form to (\ref{e:rho}) (up to re-labellings), with each state  $\rho_\mathrm{W}^{(n)}$ being a mixture of translated versions of $\rho_\mathrm{W}$.

As the weight is only shifted up or down by $\epsilon$ when the system and bath are in $ \kets{E_0}{\mathrm{S}}\kets{1}{\mathrm{B}}$ or $\kets{E_1}{\mathrm{S}}\kets{0}{\mathrm{B}}$  respectively,  the work extracted is given by
\begin{equation} \label{e:delW}
\delW =   \epsilon p_0\left(\frac{q_1}{q_0+q_1}\right) - \epsilon  p_1 \left(\frac{q_0}{q_0+q_1}\right) =  \epsilon \delp.
\end{equation}

It is straightforward to see that
\begin{equation} \label{e:delU}
\delU = \sum_k (q_k-p_k) E_k  =  \delp (E_1-E_0).
\end{equation}
The change in the entropy of the system is given by
\begin{align}
\delS &=   -q_0 \log q_0 - q_1 \log q_1 + p_0 \log p_0 + p_1 \log p_1 \nonumber \\
&= - p_0 \log \left(\frac{q_0}{p_0}\right) - p_1 \log \left(\frac{q_1}{p_1}\right)+ \delp \log \left(\frac{q_0}{q_1}\right) \nonumber \\
&= - p_0 \log \left(1 - \frac{\delp}{p_0} \right) - p_1 \log \left(1 + \frac{\delp}{p_1}\right)+\delp \frac{E_\mathrm{B}}{T} \nonumber \\
&=  \delp \frac{E_\mathrm{B}}{T} + \order{\delp^2}. \label{e:delS}
\end{align}
where in the last step we have used the fact that $\log(1+x) = x - \order{x^2}$ for $|x|<1$. Hence to first order in $\delp$
\begin{equation}
\delF \approx  \delp (E_1-E_0-E_\mathrm{B}) = - \delp \epsilon = - W.
\end{equation}
This shows that we extract work equal to the reduction in free energy of the system, up to a deficit of $\order{\delp^2}$.

Next, we consider increasing the occupation probability of an energy state which initially has probability zero. This situation would arise if we were trying to extract work from an initial pure state, as after stage 1 of our protocol the state would be $\proj{E_0}$.

Let us consider the case in which we increase $p_1$ from $0$ to $r$ in $N$ steps, whilst $p_0$ is decreased from $s$ to $s-r$. After $k$ steps, the occupation probabilities for the system will be given by
\begin{eqnarray}
p_1^{[k]} &=& k \delp \\
p_0^{[k]} &=& s - k \delp
\end{eqnarray}
where $\delp = \frac{r}{N}$. From equations (\ref{e:delU}) and (\ref{e:delW}), it follows respectively that the change in the internal energy of the system during the $k^{\textrm{th}}$ step will be
\begin{equation}
\delU^{[k]} = \delp (E_1 - E_0),
\end{equation}
and the work extracted will be
\begin{equation}
\delW^{[k]} = \delp \left(E_\mathrm{B}^{[k]} - (E_1-E_0) \right),
\end{equation}
where $E_\mathrm{B}^{[k]}$ is the energy gap of the bath qubit used in the $k^{\textrm{th}}$ step. We recall also that since we are considering thermal states this energy gap satisfies the relation
\begin{equation}
E_\mathrm{B}^{[k]}= T \log \left( \frac{p_0^{[k+1]}}{p_1^{[k+1]}}\right)
\end{equation}

It remains to calculate the entropy change of the system in the $k^{\textrm{th}}$ step,
\begin{align} \label{e:delSk}
\delS^{[k]} =&  -p_0^{[k+1]} \log p_0^{[k+1]}  -p_1^{[k+1]} \log p_1^{[k+1]} \nonumber \\ &\quad+  p_0^{[k]} \log p_0^{[k]}+ p_1^{[k]} \log p_1^{[k]} \nonumber
\end{align}
When $k=0$, this is equal to
\begin{align}
\delS^{[0]} &= - (s-\delp) \log  (s-\delp) - \delp \log \delp + s \log s \nonumber \\
&= \delp \frac{E_\mathrm{B}^{(k)}}{T} -s \log \left(\frac{s-\delp}{s}\right) \\
&= \delp \frac{E_\mathrm{B}^{(k)}}{T} + \delp + \order{\delp^2}.
\end{align}
When $k>0$, it is given by \eqref{e:delSk}
\begin{align}
\delS^{[k]} &= \delp \frac{E_\mathrm{B}^{(k)}}{T}-p_0^{[k]} \log \left( \frac{p_0^{[k+1]}}{p_0^{[k]}}\right) -p_1^{[k]} \log \left( \frac{p_1^{[k+1]}}{p_1^{[k]}}\right) \nonumber \\
&= \delp \frac{E_\mathrm{B}^{(k)}}{T}-p_0^{[k]} \log \left(1- \frac{\delp}{p_0^{[k]}}\right) -k \delp \log \left( 1+\frac{1}{k} \right)  \nonumber \\
&=\delp \frac{E_\mathrm{B}^{(k)}}{T} + \delp \left( 1- k\log \left( 1+\frac{1}{k} \right) \right)+ \order{\delp^2}.
\end{align}
By expanding the logarithm as a power series in $\frac{1}{k}$, we find
\begin{equation}
1- k\log \left( 1+\frac{1}{k} \right) = \frac{1}{2 k} - \frac{1}{3 k^2} + \frac{1}{4 k^3} - \ldots
\end{equation}
As this is an alternating sequence with  terms of decreasing magnitude it follows that
\begin{equation}
0 \leq 1- k\log \left( 1+\frac{1}{k} \right) \leq \frac{1}{2k}
\end{equation}
and hence
\begin{equation}
|\delF^{[k]} + \delW^{[k]}| \leq  T \frac{\delp}{2k} + \order{\delp^2}
\end{equation}
To obtain the total discrepancy between the work extracted and the free energy loss, we must sum over all steps $k \in \{0,1,2,\ldots,N-1\}$, obtaining
\begin{align} \label{e:0deficit}
|\delF + W | &\leq T \delp  \left( 1 + \frac{1}{2} \sum_{k=1}^{N-1} \frac{1}{k} \right) + N \order{\delp^2} \nonumber \\
&\leq T \frac{r}{N} \left( 1 +\frac{1}{2} \left( 1+ \log N\right) \right) + N \order{\frac{1}{N^2}} \nonumber \\
&=\order{\frac{\log N}{N}},
\end{align}
where in the second line we have used the fact that
\begin{equation}
\sum_{k=2}^{N} \frac{1}{k} \leq \int_1^N \frac{1}{x} \textrm{d}x =  \log N.
\end{equation}
It follows from (\ref{e:0deficit}) that as $N\rightarrow \infty$ the work extracted by the protocol approaches the free energy loss of the system, as desired.

\subsection{Carnot engine details}
Consider two thermal baths, a hot bath with temperature $T_\mathrm{H}$, and a cold bath with temperature $T_\mathrm{C} < T_\mathrm{H}$. As before, we also have a quantum system (used as a working system that links the two baths) and a weight. Imagine that the system is initially in the thermal state relative to the cold bath (with internal energy $U_\mathrm{C}$ and entropy $S_\mathrm{C}$). Our Carnot cycle is as follows: First bring the system into contact with the hot bath and use the protocol given in the Appendix to transform it into the thermal state at temperature $T_\mathrm{H}$ (with internal energy $U_\mathrm{H}$ and entropy $S_\mathrm{H}$). In the asymptotic limit, this allows us to extract work  \begin{equation} \label{e:W_i}
\qquad W_{(i)} = - \Delta F_{(i)} = (U_\mathrm{C} - T_\mathrm{H} S_\mathrm{C}) - (U_\mathrm{H}  - T_\mathrm{H} S_\mathrm{H})
\end{equation}
Second, move the system back into contact with the cold bath and use the same protocol to transform it back into the thermal state at temperature $T_\mathrm{C}$, extracting work  \begin{equation}
\qquad W_{(ii)}= - \Delta F_{(ii)} = (U_\mathrm{H} - T_\mathrm{C} S_\mathrm{H})  -(U_\mathrm{C}  - T_\mathrm{C} S_\mathrm{C}).
\end{equation}

The total work extracted in both steps is
\begin{equation} \label{e:CarnotW}
W = W_{(i)} + W_{(ii)} = (T_\mathrm{H} - T_\mathrm{C})(S_\mathrm{H} - S_\mathrm{C}).
\end{equation}
Now, by applying the first law of thermodynamics ($\Delta U = Q-W$) to the first step above, we find
\begin{equation}
 U_\mathrm{H} - U_\mathrm{C} = Q_\mathrm{H} - W_{(i)},
\end{equation}
where $Q_\mathrm{H}$ is the heat flow out of the hot bath. Substituting this in equation (\ref{e:W_i}) we obtain
\begin{equation}
 Q_\mathrm{H} = T_\mathrm{H} (S_\mathrm{H} - S_\mathrm{C})
\end{equation}
Finally, combining this with (\ref{e:CarnotW}) we find
\begin{equation}
\frac{W}{Q_\mathrm{H}}  = 1- \frac{T_\mathrm{C}}{T_\mathrm{H}}
\end{equation}

\subsection{Example of average energy conserving unitary}
Throughout the paper we treated the unitaries in an abstract way. It is instructive however to give a concrete example of how an average energy conserving unitary could be implemented. In particular, we consider here stage 1 of our protocol, as this is the only part which satisfies average energy conservation but not strict energy conservation.  

Suppose our system is a spin 1/2 particle in a magnetic field of magnitude $B$ polarized along the $z$ direction. The two energy eigenstates are 
$|\uz\ra$ and $|\dz\ra$, i.e. spin polarized ``up'' or ``down'' along the  $z$ axis, corresponding to the energy eigenvalues  $E_1=-E_2=\hbar \omega$ with $\omega={1\over2}\gamma B$ where $\gamma$ is the gyromagnetic factor. Hence $H_\mathrm{S} = \hbar \omega \sigma_z$, where $\sigma_i$ for $i\in \{x,y,z\}$ denote the usual Pauli operators. 

Let the state of the system be some arbitrary given density matrix $\rho_\mathrm{S}$. Upon diagonalisation $\rho_\mathrm{S}$ can be written as
\beq \rho_\mathrm{S}=p|\Psi_1\ra\la\Psi_1|+(1-p)|\Psi_2\ra\la\Psi_2|\eeq
with $|\Psi_1\ra$ and $|\Psi_2\ra$ being the eigenstates of $\rho_\mathrm{S}$ and $p$ a real number satisfying $0\leq p\leq 1$.  The states  $|\Psi_1\ra$ and $|\Psi_2\ra$  are orthogonal to each other (being eigenstates of $\rho_\mathrm{S}$) and will, in general be superpositions of energy eigenstates -- hence they contain coherences between energy levels. Without any loss of generality we can take $|\Psi_1\ra$ to be the state of the spin polarised in an arbitrary direction in the $x$-$z$ plane, i.e.
\beq |\Psi_1\ra=\cos{\tfrac{\theta}{2}}|\uz\ra+\sin{\tfrac{\theta}{2}}|\dz\ra\eeq
where $\theta$ is the angle it forms with the $z$ axis. The orthogonal state $|\Psi_2\ra$ is therefore 
\beq |\Psi_2\ra=-\sin{\tfrac{\theta}{2}}|\uz\ra+\cos{\tfrac{\theta}{2}}|\dz\ra.\eeq
The average energy of $|\Psi_1\ra$ is $\la\Psi_1|H_\mathrm{S}|\Psi_1\ra=\hbar\omega\cos{\theta}$ and that of   $|\Psi_2\ra$ is $\la\Psi_2|H_\mathrm{S}|\Psi_2\ra=-\hbar\omega \cos{\theta}$. 

In this particular case, the unitary for stage 1 of our protocol is given by 
\beq V=\ket{\uz} \bra{\Psi_1} \otimes \Gamma_{\epsilon} + \ket{\dz} \bra{\Psi_2} \otimes \Gamma_{-\epsilon}, \eeq
where $\epsilon=\bra{\Psi_1} H_\mathrm{S} \ket{\Psi_1} - \hbar \omega = \hbar \omega \left(\cos\theta -1\right)$. 

One straightforward way to implement this unitary would be to first apply a field which performs the rotation on the spin, and then to perform a conditional shift on the weight given the state of the spin. More concretely, we could first apply $U_1 = \ket{\uz} \bra{\Psi_1} + \ket{\dz} \bra{\Psi_2}$ followed by $U_2 = \exp(-i \epsilon \sigma_z \otimes \hat{p}/mg\hbar)$. Here, however, although the product $V = U_2 U_1$ is an interaction which preserves average energy, neither $U_1$ nor $U_2$ does individually. Although all we need is for the overall unitary $V$ to conserve the average energy, one may like a more refined protocol that conserves average energy at all times.

One can do so by moving to a continuous time picture, thus specifying an interaction Hamiltonian $H_\mathrm{int}(t)$ which implements $V$ after time $\tau$, such that if the interaction were switched off at an intermediate time $t'$, the unitary implemented would still be average energy conserving. Note that this requires us  to preserve the expected value of the free Hamiltonian $H_\mathrm{S} + H_\mathrm{W}$ at all times, rather than the full  Hamiltonian $H_\mathrm{S} + H_\mathrm{W} + H_\mathrm{int}(t) $. The latter could also be conserved if desired by adding a time-dependent constant to the Hamiltonian. We will take $\tau$ to be sufficiently short that we can neglect the free evolution of the weight during the interaction -- for larger $\tau$ the weight will also pick up some additional phases due to its free evolution, but we will nevertheless  extract the same amount of work and perform the same transformation on $\rho_\mathrm{S}$.  

Such an $H_\mathrm{int}(t)$ can be constructed by considering the simple example given above, by continuously rotating the spin and conditionally shifting the weight. More precisely, consider the interaction Hamiltonian 
\begin{equation}
H_\mathrm{int}(t) = -\hbar \omega \sigma_z - \frac{\hbar\theta}{2\tau} \sigma_y- \frac{\hbar \omega\theta}{mg\tau}\sin\left(\theta(1-\tfrac{t}{\tau})\right)\sigma(t) \otimes \hat{p}
\end{equation}
where
\begin{equation}\label{e:sig t}
\sigma(t) = \cos\left({\tfrac{\theta}{2}(1-\tfrac{t}{\tau})}\right) \sigma_z + \sin\left({\tfrac{\theta}{2}(1-\tfrac{t}{\tau})}\right)\sigma_x.
\end{equation}
Note that this interaction Hamiltonian is transnationally invariant on the weight, as we would desire in our formalism. If the last term of \eqref{e:sig t} were excluded it is straightforward to see that the effect of $H_\mathrm{S}+H_{int}(t)$ would be to  rotate the system spin into the energy eigenbasis of $H_\mathrm{S}$ in time $\tau$, with the initial eigenstates of $\rho_\mathrm{S}$ tranforming at time $t$ into
\begin{eqnarray}
	|\Psi_1(t)\ra&=&\cos\left({\tfrac{\theta}{2}(1-\tfrac{t}{\tau})}\right)|\uz\ra+\sin\left({\tfrac{\theta}{2}(1-\tfrac{t}{\tau})}\right)|\dz\ra \\
	|\Psi_2(t)\ra&=&-\sin\left({\tfrac{\theta}{2}(1-\tfrac{t}{\tau})}\right)|\uz\ra+\cos\left({\tfrac{\theta}{2}(1-\tfrac{t}{\tau})}\right)|\dz\ra\nonumber
\end{eqnarray}
These states are instantaneous eigenstates of $\sigma(t)$, hence the last term in the interaction Hamiltonian does not affect the evolution of $\rho_\mathrm{S}$.

However, on the weight this additional term now produces the desired conditional shift, conditioned on the instantaneous eigenstates of $\sigma(t)$. The rate at which we need to move the weight is given by the rate of change in the average energy of the system, 
\begin{equation}
	\frac{d}{dt}\bra{\Psi_i(t)} H_\mathrm{S} \ket{\Psi_i(t)} = \pm\frac{\hbar \omega\theta}{\tau}\sin\left(\theta(1-\tfrac{t}{\tau})\right)
\end{equation}  

This thus constitutes a model that will implement the desired evolution whilst conserving the average energy  throughout the interaction time $\tau$. Note that the same evolution would work for any state with the same eigenbasis as $\rho_\mathrm{S}$.

\subsection{Limitations of protocols satisfying strict energy conservation}

Here we consider protocols satisfying strict energy conservation (i.e. unitaries that commute with the total Hamiltonian).  We will show that such protocols cannot extract work equal to the full change in free energy for systems in initial states having coherences between energy levels, following a similar approach to \cite{HorOpp11}.

Firstly, note that as we require the average work extracted by a protocol to be independent of the initial state of the weight, we are free to choose that state however we like -- here we choose it to be a very narrow wavepacket centred on zero. Now consider a decomposition of the total state space into subspaces, each of which has total energy (of the system, bath and weight) close to one of the energy eigenvalues of the system and bath, in particular the $i^{\textrm{th}}$ subspace  corresponds to the total energy $E$ lying in the range $\frac{E_i-E_{i-1}}{2} \leq E \leq \frac{E_{i+1}-E_{i}}{2}$, where $E_i$ are the energy eigenvalues of the system and bath. Furthermore we choose the width of the weight's initial wavepacket to be narrower than the smallest subspace. Note that any work extraction protocol can be followed by a transformation which decoheres the state with respect to these total energy subspaces, without affecting the average work extracted. However, this decohering operation commutes with the unitaries used in the protocol, so we can move it to the beginning of the protocol without changing the work extracted. At the beginning of the protocol, this operation has the sole effect of decohering the system in its energy eigenbasis (changing $\rho$ to $\omega=\sum_i\Pi_i\rho\Pi_i$,  where $\Pi_i$  is the projector  onto the $i$th energy subspace ). Hence the protocol extracts the same amount of work as it would have if it had operated on $\omega$ and therefore there is a work “deficit” equal to $F(\rho)-F(\omega)$.

\end{document}